# Amino acid characteristics in protein native state structures

Tatjana Škrbić[1,2], Achille Giacometti[1,3], Trinh X. Hoang[4], Amos Maritan[5], and

Jayanth R. Banavar[2]

[1] Ca' Foscari University of Venice, Department of Molecular Sciences and

Nanosystems, Venice, Italy

[2] University of Oregon, Department of Physics and Institute for Fundamental

Science, Eugene, Oregon, USA

[3] European Centre for Living Technology (ECLT), Ca' Bottacin, Dorsoduro

3911, Calle Crosera, 30123 Venice, Italy

[4] Institute of Physics, Vietnam Academy of Science and Technology, 10 Dao

Tan, Ba Dinh, Hanoi 11108, Vietnam

[5] University of Padua, Department of Physics and Astronomy, Padua, Italy

**Correspondence**

Tatjana Škrbić

Ca' Foscari University of Venice, Department of Molecular Sciences and

Nanosystems, Via Torino 155, 30170 Venice, Italy.

Email: tatjana.skrbic@unive.it

**Abstract**

We present a geometrical analysis of the protrusion statistics of side chains in more than 4,000 high-resolution protein structures. We employ a coarse-grained representation of the protein backbone viewed as a linear chain of $C_\alpha$ atoms and consider just the heavy atoms of the side chains. We study the large variety of behaviors of the amino acids based on both rudimentary structural chemistry as well as geometry. Our geometrical analysis uses a backbone Frenet coordinate system for the common study of all amino acids. Our analysis underscores the richness of the repertoire of amino acids that is available to nature to design protein sequences that fit within the putative native state folds.

*Keywords:* local Frenet frame, amino-acid classes, side-chain protrusion, pre-sculpted landscape

# I. INTRODUCTION

Proteins, the molecular machines of life, are relatively short linear chains of amino acids with a common backbone. There are twenty types of naturally occurring amino-acids, each possessing a distinct side chain attached to the main chain protein backbone [1-4]. The complexity of the protein problem stems from the myriad of degrees of freedom. A protein is surrounded by water molecules within the cell. Each of the twenty side chains has its own chemical properties and geometry. Despite the complexity, small globular proteins share a great deal of properties because of their common backbone. They fold rapidly and reproducibly into their respective unique native state structures [5]. Protein native state structures are modular and comprise secondary structure building blocks: topologically one-dimensional α-helices and almost planar parallel and antiparallel β-sheets. Hydrogen bonds provide support to the building blocks [6,7]. Because of the modularity, the total number of distinct native folds is of the order of just several thousand [8-11]. Furthermore, the native state folds are evolutionarily conserved [12,13]. This surprising simplicity present in the complex protein problem can be rationalized through the notion of a free energy landscape of proteins sculpted by the common backbone of all proteins [14-21].

The side chains play a critical role in the selection process in two crucial ways. First, the chemistry of the interacting side chains [22-25] must be harmonious [26-28], maximizing favorable interactions (including water-mediated hydrophobic, van der Waals, electrostatic, and hydrogen bonding interactions). The net result is to create a protein hydrophobic core shielded from the surrounding water molecules thereby ensuring the stability and compactness of protein native structure. Second, the side chains must fill the space in the interior of the protein, packing tightly against each other, maximizing favorable self-interactions in the hydrophobic interior, and minimizing empty space [29] (see Figure 1). Interestingly, even in toy chain models [32-38], adding side chain spheres to the canonical tangent sphere model and permitting adjoining spheres to overlap, destabilizes the disordered compact globular phase and results in novel structured phases with effectively reduced dimensionalities.

The specific arrangement of side chains within the protein interior is determined by at least two factors. The first is the primary protein sequence of amino acids that can grossly be classified as being hydrophobic (non-polar residues mainly buried in the protein interior and forming its hydrophobic core), hydrophilic (polar or charged residues that readily interact with water molecules and tend to be positioned at the protein surface), or neutral (somewhere between the two categories) [39]. The second is that the overall folded geometry ought to provide an optimal, best possible fit to the sequence. The orientation of the side chain is flexible and the set of specific conformations and/or orientations that are statistically significant constitute the so-called side chain rotamers [40-44]. There could also be an entropic cost associated with freezing a side chain into a particular rotamer conformation, which may be more relevant in the denatured state.

Here we adopt a simplified coarse-grained description. We view a protein as a chain of $C_\alpha$ atoms. Our approach then consists in the determining the locations and orientations of the protruding side chain atoms. Because of the imperative need to fill space in the interior while assiduously avoiding steric clashes, our focus is on the heavy atom protruding furthest from the corresponding $C_\alpha$ atom. We seek to characterize the geometry of this protrusion in a universal coordinate frame relative to the portion of protein backbone corresponding to the given amino acid, which enables us to determine both the average side chain behavior as well as the specific behavior of distinct amino acids. Our longer-term goal is to set the stage for the more important step of understanding the role of side chains in tertiary structure assembly. This will enable us to understand how a particular sequence chooses a specific fold geometry for its native state.

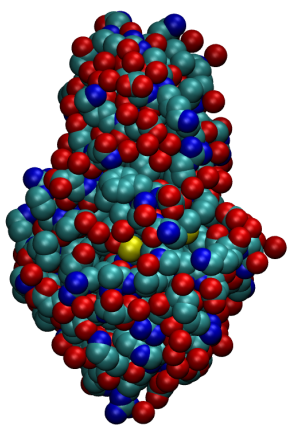

Figure 1: a) Native state of bacteriophage T4 lysozyme (PDB code: 2LZM) in the CPK representation [30-31] in which all heavy atoms of the protein backbone and its side chains are represented as spheres with radii proportional to their respective van der Waals atomic radii. Color code: carbon (cyan), oxygen (red), nitrogen (blue), and sulfur (yellow). The side chains in the protein interior are very well packed.

## II. MATERIALS AND METHODS

### A. Local Frenet coordinate system of an amino acid

We view a protein backbone as a chain of discrete points on which the consecutive $C_\alpha$ atoms are located. For the side chains, we account for all heavy atoms (excluding only the hydrogen atoms) to determine the maximally protruding atom from the protein backbone. This is the farthest heavy atom from the corresponding $C_\alpha$ atom and at a distance that we call $R_{max}$. To characterize the orientation of this maximally protruding side chain atom, we employ a Frenet coordinate system [45] local to the portion of the backbone to which the side chain belongs. For the i-th amino acid in question, the origin of its local Frenet frame is located at the i-th $C_\alpha$ atom. The orthonormal set of axes are the tangent **t**, anti-normal **an**=-**n**, and binormal **b**. These basis vectors are defined from the positions of three consecutive $C_\alpha$ atoms associated with residues i-1, i, and i+1, as shown in Figure 2.

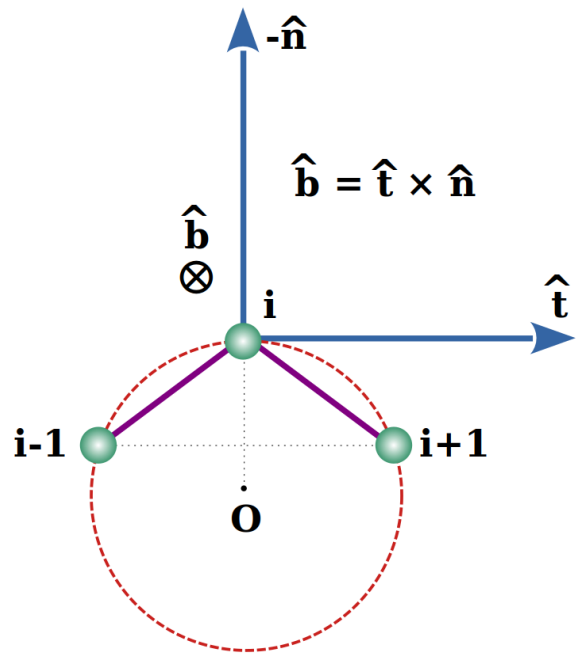

Figure 2: Local Frenet frame of the amino acid i. The three consecutive $C_\alpha$ atoms are at points i-1, i, and i+1 and lie in the plane of the paper. The point O is at the center of a circle passing through them. Please see text for a description of the orthonormal basis set.

99.7% of the bond lengths in proteins are, to a very good approximation, equal to 3.81Å [46], corresponding to the prevalent *trans* isomeric conformation of a peptide backbone group, where the two neighboring $C_\alpha$ atoms along the chain are on opposite sides of the peptide bond with the third Ramachandran angle of ω close to 180°. However, the remaining ~0.3% of protein bonds are shorter, having a length around ~2.95Å [46] and correspond to the so-called *cis* conformation of a backbone [47], in which the two consecutive $C_\alpha$ atoms are placed on the same side of the connecting peptide bond, when the third Ramachandran angle is ω ~ 0°. We define a local Frenet frame of a given amino acid in a manner that is robust to variations in the bond lengths. First, independent of the bond lengths, we draw a circle passing through points i-1, i, and i+1 and determine its center and the radius. The direction of the anti-normal (negative normal direction) **an**=**-n** is along the straight line joining the center of the circle to the $C_\alpha$ atom. The tangent vector **t** points along the direction (i-1, i+1). Both the tangent and normal vectors are in the plane of the paper in Figure 2. The binormal vector **b** is found as a cross product of the unit vectors **t** x **n** and is perpendicular and into the plane of the paper (see Figure 2). The Frenet frame is well defined at all but the end sites of a protein chain and serves as a convenient reference frame for studying the side chain protrusion of all amino acids in the native state structures. We characterize the orientation of the maximally protruding heavy atom of the side chain from the $C_\alpha$ atom by means of three projections, along the unit vectors **t**, **b**, and **-n**, in the corresponding local Frenet system.

**B. Curation and data analysis**

Our protein data set consists of 4,366 globular protein structures from the PDB, a subset of Richardsons' Top 8000 set [48] of high-resolution, quality-filtered protein chains (resolution < 2Å, 70% PDB homology level), that we further distilled out to exclude structures with missing backbone and side chain atoms, as well as amyloid-like structures. The program DSSP (CMBI version 2.0) [49] has been used to determine for each protein residue its context: in an α-helix, in a β-strand or elsewhere.

Our data set comprises of a total of 959,691 residues (883,407 non-glycine and 76,284 glycine amino acids) in the native state structures of more than 4,000 proteins. Their abundances and relative frequencies, in order of decreasing prevalence, in our data set, are shown in Table I.

| Type | Total number | Frequency [%] | α | β | loop |
|---|---|---|---|---|---|
| **LEU** | 84,916 | 8.85 | 36,154 | 21,387 | 27,375 |
| **ALA** | 82,208 | 8.57 | 38,896 | 13,583 | 29,729 |
| **GLY** | 76,284 | 7.95 | 10,839 | 10,883 | 54,562 |
| **VAL** | 69,481 | 7.24 | 20,194 | 29,569 | 19,718 |
| **GLU** | 61,780 | 6.44 | 28,135 | 9,678 | 23,967 |
| **ASP** | 57,111 | 5.95 | 15,259 | 6,795 | 35,057 |
| **SER** | 56,318 | 5.87 | 13,965 | 10,649 | 31,704 |
| **ILE** | 54,043 | 5.63 | 18,561 | 20,635 | 14,847 |
| **LYS** | 53,739 | 5.60 | 20,349 | 9,605 | 23,785 |
| **THR** | 53,588 | 5.58 | 13,129 | 14,272 | 26,187 |
| **ARG** | 46,176 | 4.81 | 18,251 | 9,217 | 18,708 |
| **PRO** | 44,397 | 4.63 | 6,396 | 4,148 | 33,853 |
| **ASN** | 42,128 | 4.39 | 9,757 | 5,804 | 26,567 |
| **PHE** | 38,853 | 4.05 | 12,348 | 12,184 | 14,321 |
| **TYR** | 34,685 | 3.61 | 10,506 | 10,825 | 13,354 |
| **GLN** | 34,361 | 3.58 | 14,372 | 5,870 | 14,119 |
| **HIS** | 22,392 | 2.33 | 6,261 | 4,897 | 11,234 |
| **MET** | 19,524 | 2.03 | 8,273 | 4,513 | 6,738 |
| **TRP** | 14,579 | 1.52 | 4,698 | 4,205 | 5,676 |
| **CYS** | 13,128 | 1.37 | 3,469 | 3,656 | 6,003 |

Table I: Total number and relative frequency of twenty amino acid types in our data set comprised of over 4,000 protein native state

structures, shown from the most abundant leucine (LEU) to the least abundant cysteine (CYS), along with the number of twenty amino acids in different protein contexts: helical 'α', strand 'β', and 'loop'.

## III. RESULTS AND DISCUSSION

### A. The orientation of amino acids in globular proteins

For each amino acid in our data set of proteins, we determine a protrusion vector in the Frenet frame which connects a $C_\alpha$ atom to the maximally protruding heavy atom in its side chain. By maximally protruding, we mean the heavy atom that is the farthest away from the $C_\alpha$ atom. This provides a rough idea of the spatial extent and the relevant direction of the side chain of the residue. The presence of rotamers in the native structures of proteins immediately implies that not all amino acids of a given type will have the same protrusion vector. Our analysis aims to determine the statistics of protrusion of all side chains and of the side chains of individual amino acid types.

Our results are summarized in Table II. We begin by averaging the protrusion vectors of all amino acids in our data set to determine an average protrusion vector, characterized by its magnitude, and the components of the normalized unit average protrusion vector along the three Frenet axes (the squares of these components add up to 1). With the notable exception of proline, the average protrusion vector lies predominantly in the (anti-normal-binormal) plane with a relatively small component in the tangent direction (see Table II). More specifically, the resulting protrusion vector averaged over all amino acids in our data set forms angles of 26.71°, 92.44°, and 116.58°, with the anti-normal, tangent and binormal vectors, respectively. Interestingly, amino acids predominantly point close to the anti-normal direction, thus avoiding the protein backbone. Additionally, the magnitude of the mean protrusion vector of all amino acids is found to be 3.81Å matching the distance between consecutive $C_\alpha$ atoms along the chain. This equality of two characteristic lengths in proteins, one along the protein backbone and the second approximately perpendicular to it, is noteworthy. Table II also presents analogous data for the nineteen amino acids possessing heavy atoms in their side chains. This excludes glycine, which has none.

To obtain a measure of the spread of the data around the average value for a given amino acid, we use two measures. The first is a ratio of the magnitude of the average protrusion vector to the average protrusion distance (measured with no regard to the varying directions), which we denote as $R_{eff}/\langle R_{max} \rangle$. We also take an average of the dot product of the individual protrusion vectors with the average protrusion vector for each amino acid and denote it as $\langle \cos \theta \rangle$. Note that the two independent estimates of the spread defined in this way are in excellent accord with each other. We note that the largest spread is displayed by amino acids with a ring structure (HIS, PHE, TRP, and TYR), followed by long linear chains (ARG, GLN, GLU, and LYS). For the gallery of the nineteen amino acid types see Figure 3.

Figure 3 depicts the vectors of the mean protrusion of the nineteen amino acids in the local Frenet frame. The magnitude of the vector is $R_{eff}$. The figure depicts three two-dimensional views. The protrusion of the side chains is dominantly in the negative binormal-negative normal plane. Even a cursory look at Figure 3 shows that PRO (gray, almost horizontal arrow in a) and b)) is an outlier. PRO has a large projection in the tangent direction (that is along the backbone direction) due to its peculiar geometry that reaches back to the protein backbone. Leaving aside Proline, we note (see Figure 3a and Figure 3b) that the projection along the anti-normal direction spans the range of 3.5 Å between 1.1 Å (ALA, red) and 4.6 Å (ARG, dark blue). For the binormal, the values range over a smaller interval from -2.2 Å (ARG, dark blue arrow) to -0.6 Å (TRP, green arrow). Finally, along the tangent (see Figures 3b and 3c), the values of the projections range from -0.7 Å (ILE, again red) to 0.5 Å (VAL, another red). Let us take a closer look at Figure 3a and the directions in which the mean vectors for a given amino acid type protrude in this plane.

    Figure 3 shows that, after PRO, ALA (red) is the next outlier. ALA with only one $C_\beta$ carbon atom in its side chain, bonded directly to $C_\alpha$ atom, has a highly constrained geometry of protrusion due to $sp^3$ hybridization of the $C_\alpha$ atom. ALA is followed by ASP (orange) and ASN (purple) sharing essentially the same geometry. Figure 4 shows that they share the same geometrical shape, the difference being that one oxygen atom in ASP is converted to nitrogen in the case of ASN. On the other side in Figure 3a, the aromatic trio, PHE (dark green), TYR (light green) and TRP (green) form the largest angles with the binormal direction (and the smallest angles with the anti-normal

direction), while sharing very similar directions. They are thus, among all amino acids, on average, pointing the most away from the backbone. On the other hand, TRP is unique in that it has a 'double ring' for its side chain (see Figure 4), and this makes its full protrusion geometry quite distinct (see Section III.C).

We have also studied the variations of Figure 3 within an individual amino acid (not shown). The striking result is that, in terms of the direction of protrusion (not magnitude), three pairs of geometrical twins show similar behaviors within a pair: (ASN and ASP); (GLN and GLU); and (PHE and TYR). Even in cases when the mean poking for an amino acid in the tangent direction is small, there are large fluctuations especially when the side chains are large in size (PHE, TRP, TYR rings and ARG, LYS linear topology).

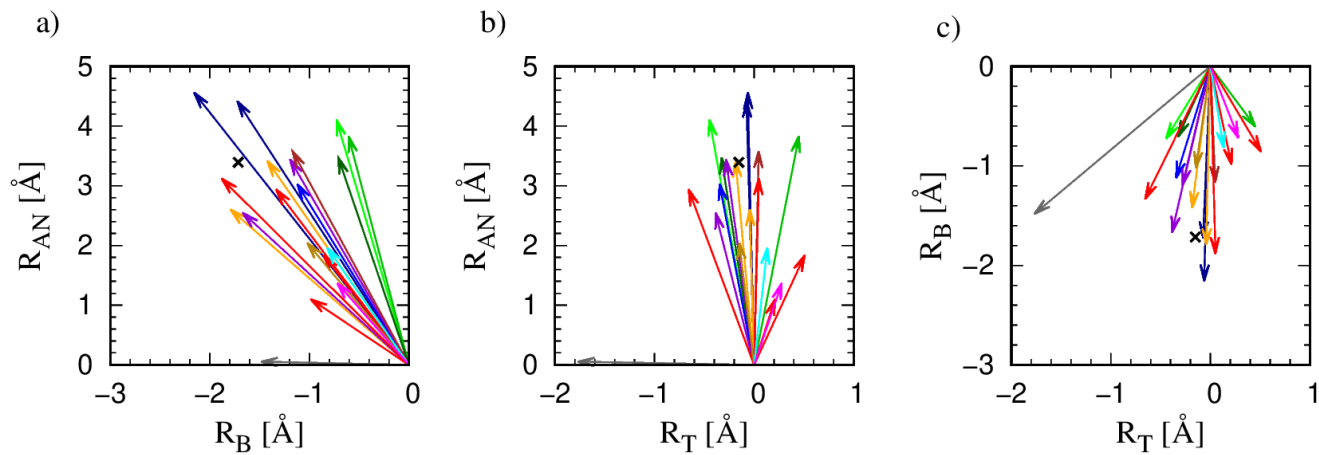

Figure 3: Two dimensional projections of the mean maximal protrusion of nineteen amino acids in more than 4,000 high-resolution structures of globular proteins. For ease of visualization, we show three two-dimensional views: a) in the anti-normal-binormal plane; b) in the anti-normal-tangent plane; and c) in the binormal-tangent plane. The color code of the protrusion vectors follows that employed in Table III. The black X symbols in all the three panels denote the end point of the projection of the mean protrusion vector calculated for *all* amino acids in our data set into the corresponding plane.

| Type | $\langle R_{max} \rangle$ [Å] | $R_{eff}$ [Å] | $R_{eff}/\langle R_{max} \rangle$ | $u_{AN}$ | $u_T$ | $u_B$ | $\langle \cos \theta \rangle$ |
|---|---|---|---|---|---|---|---|
| **All** | **3.81** | **2.99** | **0.78** | **0.89** | **-0.04** | **-0.45** | **0.79** |
| **PRO** | 2.43 | 2.31 | 0.95 | 0.02 | -0.77 | -0.64 | 0.95 |
| **ALA** | 1.53 | 1.49 | 0.98 | 0.74 | 0.14 | -0.66 | 0.98 |
| **ILE** | 3.73 | 3.29 | 0.88 | 0.89 | -0.20 | -0.41 | 0.88 |
| **LEU** | 3.90 | 3.64 | 0.94 | 0.86 | 0.01 | -0.52 | 0.93 |
| **VAL** | 2.54 | 2.09 | 0.82 | 0.88 | 0.24 | -0.41 | 0.82 |
| **PHE** | 5.12 | 3.56 | 0.69 | 0.98 | -0.09 | -0.20 | 0.69 |
| **TRP** | 6.13 | 3.90 | 0.64 | 0.98 | 0.11 | -0.16 | 0.63 |
| **TYR** | 6.45 | 4.19 | 0.65 | 0.98 | -0.11 | -0.17 | 0.65 |
| **ARG** | 6.49 | 5.04 | 0.78 | 0.90 | -0.01 | -0.43 | 0.77 |
| **LYS** | 5.78 | 4.74 | 0.82 | 0.93 | -0.02 | -0.36 | 0.81 |
| **HIS** | 4.56 | 3.24 | 0.71 | 0.93 | -0.11 | -0.34 | 0.71 |
| **ASP** | 3.61 | 3.16 | 0.87 | 0.82 | -0.01 | -0.57 | 0.87 |
| **GLU** | 4.60 | 3.71 | 0.81 | 0.92 | -0.05 | -0.38 | 0.80 |
| **ASN** | 3.60 | 3.07 | 0.85 | 0.83 | -0.13 | -0.54 | 0.85 |
| **GLN** | 4.54 | 3.65 | 0.80 | 0.94 | -0.08 | -0.33 | 0.79 |
| **SER** | 2.43 | 2.00 | 0.83 | 0.68 | 0.14 | -0.72 | 0.83 |
| **THR** | 2.53 | 2.13 | 0.84 | 0.92 | 0.06 | -0.38 | 0.84 |
| **CYS** | 2.80 | 2.29 | 0.82 | 0.89 | -0.06 | -0.45 | 0.82 |
| **MET** | 4.54 | 3.76 | 0.83 | 0.95 | 0.01 | -0.31 | 0.82 |
| **GLY** | - | - | - | - | - | - | - |

Table II. Statistics of the protrusion for all amino acids in our data set, as well as for the nineteen amino acids separately.

**B. The protruder atom type and amino acid groupings**

Figure 4 indicates, for each of the nineteen amino acids (but glycine (GLY) that does not possess any heavy atoms), the atom that protrudes the most along with the percentage of time it does. We note that in most cases there is prevalently only one such atom (~90% or more) and this is the case for the thirteen amino acids: ALA, ASN, ASP, CYS, ILE, LYS, MET, SER, THR, TYR, TRP, PHE, and PRO. For the remaining six amino acids: ARG, GLN, GLU, HIS, LEU, and VAL there were two viable candidate atoms. We note that both hydrophilic and hydrophobic residues are present in both these classes showing that this result is largely chemistry independent.

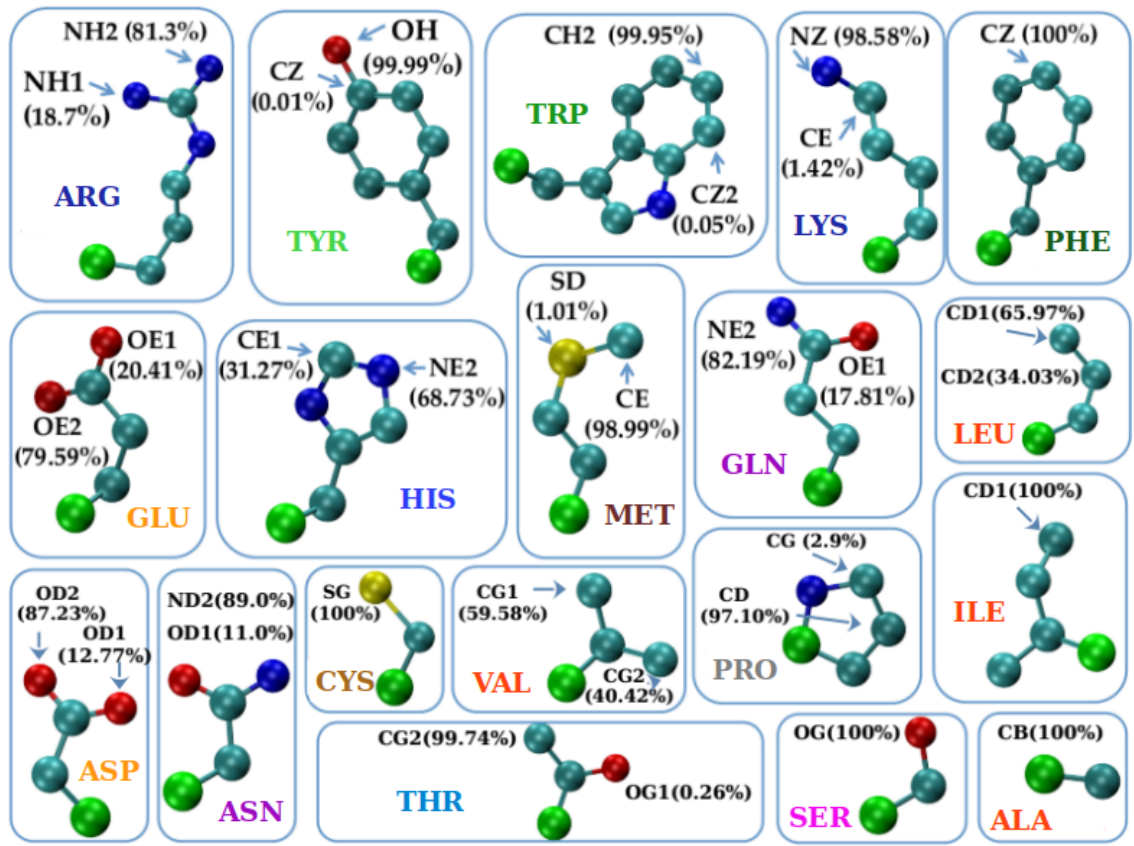

Figure 4: Gallery of nineteen amino acids (with glycine excluded). Three-letter amino acid codes are used. For each amino acid, the maximally protruding atom along with the frequency with which it occurs is shown. The color code of the atoms is: carbon $C_\alpha$ in green, carbon C atoms other than $C_\alpha$ in turquoise, oxygen O atoms in red, nitrogen N atoms in dark blue, and sulfur S atoms in yellow. Carbon $C_\alpha$ atoms (green spheres) are artificially represented as spheres with slightly larger radius than the rest of C atoms (cyan spheres) to enhance visibility. The measure of the degree of protrusion of a given side chain atom with respect to the backbone was defined to be the distance of the atom from the corresponding $C_\alpha$ atom. The color code of the amino acid labels follows that in Table III.

Based on Figure 4, we now proceed to a coarse graining of the amino acids into similar groups. The combination of rudimentary structural chemistry and protrusion geometry allows us to crudely divide our amino acids into 14 groups. Glycine is a group unto itself because it has not side chain heavy atoms. Likewise, proline is special because it has a ring that connects back to the backbone. The rest of the amino acids can be grouped together based on the topology of the side chain (linear or ring) and the identities of the non-carbon atoms in the side chain and the most protruding one. This yields one group with 4 amino acids and two groups with 2 amino acids each and twelve singlet groups in all. Interestingly, the 11 groups in the IMGT classification [50] result from a partial merger of our 14 groups: Group VI (ARG, LYS) with VII (HIS); Group X (SER) with XI (THR); and Group XII (CYS) with XIII (MET). Amino acids (ARG, LYS, HIS) form the so-called 'basic' IMGT group, composed of all positively charged amino acids among the nineteen, while (SER, THR) constitute the 'hydroxylic' IMGT group of polar amino acids that contain the -OH group. Finally, (CYS, MET) form the so-called 'sulfur-containing' IMGT group, as the only two amino acids that contain a sulfur atom. We now turn to a careful analysis of the geometry of protrusion of the side chains.

| *Group I* | *PRO* | *Ring connects back to the backbone* |
|---|---|---|
| *Group II* | *ALA, ILE, LEU, VAL* | *Linear (C); C: max* |
| *Group III* | *PHE* | *Ring (C); C: max* |
| *Group IV* | *TRP* | *Ring (C, N); C: max* |
| *Group V* | *TYR* | *Ring (C, O); O: max* |
| *Group VI* | *ARG, LYS* | *Linear (C, N); N: max* |
| *Group VII* | *HIS* | *Ring (C, N); N: max* |
| *Group VIII* | *ASP, GLU* | *Linear (C, O, O); O: max* |
| *Group IX* | *ASN, GLN* | *Linear (C, N, O); N: max* |
| *Group X* | *SER* | *Linear (C, O); O: max* |
| *Group XI* | *THR* | *Linear (C, O); C: max* |
| *Group XII* | *CYS* | *Linear (C, S); S: max* |
| *Group XIII* | *MET* | *Linear (C, S); C: max* |
| *Group XIV* | *GLY* | *No heavy atoms* |

Table III. Classification of amino acids into 14 groups, based on the side chain topology, the type of atoms it contains, and the type of the atom that is maximally protruding from the corresponding $C_\alpha$ atom.

## C. The geometry of amino acid protrusion

We have observed that the mean protrusion vector calculated over all amino acids lies predominantly in the anti-normal-binormal plane of the corresponding local Frenet frames (see also Table II). This information allows us to considerably simplify our analysis and concentrate on the protrusion behavior in this plane. To this end, we define ε as the angle made by the projection of an individual amino acid in the anti-normal-binormal plane with the anti-normal direction. For each of the nineteen amino acids (except for glycine, which has no heavy atoms in its side chain), we measure the distribution of ε. The mean, the modal value(s) (there are sometimes multiple modes), and the standard deviations are shown in Table IV. We have carried out the calculations based on the context (helix α, strand β, or loop) of the amino acids. The lessons learned are the following:

- PRO due to its distinct geometry of a ring that reconnects to the protein backbone, has characteristic ε values that are close to or even larger than 90°. This context-independent result reflects the fact that PRO dominantly protrudes in the binormal-tangent plane unlike all the other amino acids (see also Table II). PRO forms the singlet 'neutral aliphatic' group in the IMGT classification [50] and is our singlet Group I (see Table III).

- ALA, ILE, LEU and VAL have qualitatively similar behaviors. For both α and β contexts, one mode strongly dominates, while in the loop context, the behavior is a combination of the modes in the α and β contexts. (ALA, ILE, LEU, VAL) form the 'hydrophobic aliphatic' IMGT group [50] and coincides with our Group II (see Table III).

- PHE and TYR share very similar behavior, with only one mode present in each of the contexts and all of them ~0°, meaning that these amino acids with aromatic rings protrude predominantly along the anti-normal direction. PHE is a singlet 'hydrophobic, aromatic, with no hydrogen donor' and TYR a singlet 'neutral, aromatic, with both hydrogen donor and acceptor' group in the IMGT classification [50], We denote them as singlet groups as well, Group III and Group V (see Table III).

- TRP is the unique amino acid with the 'double ring' structure (composed of a six-atom ring and a five-atom ring, sharing one side, see Figure 3) and, contrary to all other amino acids has an ε angle α-mode smaller than the ε angle β-mode. TRP forms the singlet 'hydrophobic, aromatic, with hydrogen donor' IMGT group [50] and is our singlet Group IV (see Table III).

- ARG, LYS, and HIS, the three positively charged amino acids forming the 'basic' group in IMGT classification [50]. They all exhibit a ~0° β-mode, but quite different α-modes. For ARG, there are two α-modes, presumably due to the 'double tip' branch formed by two symmetrically placed nitrogen atoms at its end (see Figure 3). In our classification, ARG and LYS fall into Group VI, while HIS forms the singlet Group VII, due to its different topology (see Table III).

- ASP and ASN, on one hand, and GLU and GLN, on the other, have very similar ε angle profiles, so they can be dubbed geometrical twins. From Figure 3, we see that this is due to the identical shape for the two corresponding pairs, with the difference that for ASP and GLU the 'double tip' in the amino acid ending is made up of two oxygen atoms, while for the ASN and GLN the double tip is composed of one oxygen and one nitrogen atom. In the IMGT categorization [50], ASP and GLU constitute the 'acidic' group, while ASN and GLN form the 'amide' group. In our classification, these pairs of amino acids form Group VIII and Group IX, respectively (see Table III).

- SER and THR constitute the 'hydroxylic' group in the IMGT classification [50] and have decisively different protrusion geometries, with SER most notably (and distinctively from all other amino acids) displaying the most complex $\varepsilon$ profile, with three $\alpha$-modes, two $\beta$-modes, as well as two loop-modes. SER is thus the champion of versatility with multiple sharp modes in all environments that is surprising because of its relatively small size. 60% of the time, SER is found in loops. In our grouping, SER and THR form two singlet groups, Group X and Group XI, respectively (see Table III).

- CYS and MET, placed in the 'sulfur-containing' group in the IMGT classification [50], have different protrusion geometries. SER has a non-zero $\alpha$-mode and zero $\beta$- and loop-modes; while MET with all three zero-modes, seems more compatible geometry-wise with the aromatic duo

PHE and TYR. In our grouping, CYS and MET are in two singlet groups, Group XII and Group XIII (see Table III).

- There are three amino acids, ARG, GLN and GLU with two dominant α-modes, that could be due to their considerable length and the 'double tip' shape in the amino acid ending. For GLN, this is also reflected in the double peak in the distribution of the magnitude of the maximal protrusion $R_{max}$ (see Figure 5), while for ARG, $R_{max}$ has a very broad distribution, so that no well-defined peaks could be identified.

- Finally, GLY (with no heavy side chain atoms) is our singlet Group XIV and it belongs to the 'very small, neutral aliphatic' singlet group in the IMGT classification [50].

| Type | $\varepsilon_\alpha$ mode [°] | $\varepsilon_\alpha$ mean [°] | $\varepsilon_\beta$ mode [°] | $\varepsilon_\beta$ mean [°] | $\varepsilon_{loop}$ mode [°] | $\varepsilon_{loop}$ mean [°] |
|---|---|---|---|---|---|---|
| **PRO** | 105 | 104.9 ± 5.5 | 77 | 74.8 ± 13.1 | 73, 108 | 83.2 ± 21.1 |
| **ALA** | 50 | 50.0 ± 2.3 | 25 | 28.2 ± 7.3 | 30, 48 | 37.7 ± 10.4 |
| **ILE** | 45 | 37.2 ±15.9 | 12 | 20.0 ± 14.6 | 12, 53 | 29.3 ± 20.3 |
| **LEU** | 43 | 40.8 ± 5.8 | 16 | 19.4 ± 9.7 | 18, 38 | 27.9 ± 12.3 |
| **VAL** | 24 | 32.7 ± 15.5 | 5 | 16.3 ± 20.7 | 7, 23 | 29.8 ± 26.3 |
| **PHE** | 3 | 14.1 ± 14.6 | 3 | 24.5 ± 28.8 | 3 | 21.1 ± 25.2 |
| **TRP** | 18 | 30.5 ± 24.4 | 32 | 36.7 ± 23.4 | 30 | 39.9 ± 30.2 |
| **TYR** | 0 | 14.1 ± 17.1 | 4 | 25.6 ± 28.6 | 4 | 24.1 ± 27.6 |
| **ARG** | 30, 70 | 38.6 ± 24.1 | 2 | 23.9 ± 20.5 | 3 | 29.9 ± 23.5 |
| **LYS** | 38 | 31.1 ± 16.5 | 7 | 20.3 ± 16.2 | 12 | 27.8 ± 19.8 |
| **HIS** | 14 | 24.8 ± 18.4 | 5 | 19.6 ± 24.3 | 0 | 26.9 ± 27.7 |
| **ASP** | 42 | 42.2 ± 10.5 | 14 | 19.8 ± 16.2 | 10, 40, 60 | 34.9 ± 21.0 |
| **GLU** | 3, 35 | 29.3 ± 17.4 | 0 | 18.8 ± 17.4 | 3 | 29.9 ± 23.3 |
| **ASN** | 43 | 40.4 ± 11.0 | 15 | 22.3 ± 16.5 | 18, 37, 57 | 34.5 ± 19.5 |
| **GLN** | 0, 29 | 28.4 ± 17.1 | 0 | 20.7 ± 17.6 | 0 | 27.7 ± 20.9 |
| **SER** | 25, 38, 78 | 49.6 ± 22.9 | 3, 58 | 30.5 ± 27.0 | 10, 77 | 48.6 ± 28.3 |
| **THR** | 23 | 26.9 ± 9.3 | 5 | 16.4 ± 20.2 | 17 | 24.6 ± 16.8 |
| **CYS** | 32 | 32.9 ± 13.6 | 0 | 18.7 ± 24.8 | 3 | 29.2 ± 27.0 |
| **MET** | 0 | 28.7 ± 21.1 | 0 | 24.9 ± 17.7 | 0 | 24.5 ± 19.3 |

Table IV. Statistics of values of the angle ε between the projection of the most protruding vector in the anti-normal-binormal plane with the anti-normal direction. The positions of the most frequently observed value (mode) or modes (when there are more than one mode) are presented. The mean values and standard deviations of the angles $\varepsilon_\alpha$,

$\varepsilon_\beta$, and $\varepsilon_{loop}$ characterizing the geometry of protrusion in three different contexts: α, β, and loop are also presented.

Finally, we have studied the distribution of the values of the maximal protrusion $R_{max}$ for each of the 19 amino acids shown in Figure 5. The observed peaks in this distribution can be readily assigned to specific amino acids because of their non-overlapping mean values and their relatively sharp widths. Additionally, we can conveniently divide the observed range of $R_{max}$ into three distinct classes: 1) small with $R_{max} < 3$Å, comprised of ALA, CYS, PRO, SER, and VAL; 2) medium $R_{max} \sim (3-5)$ Å, composed of ASN, ASP, GLN, GLU, HIS, ILE, LEU, and MET; and 3) large with $R_{max} > 5$Å, containing ARG, LYS, PHE, TRP, and TYR.

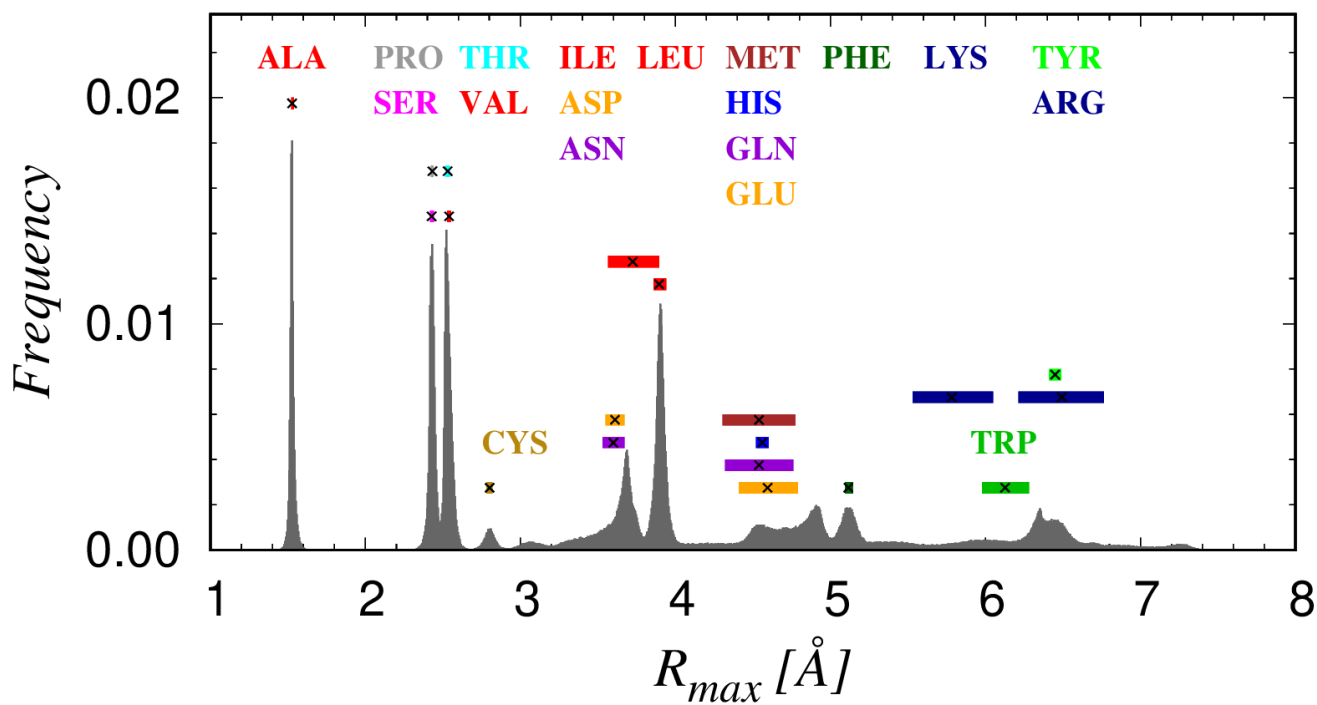

Figure 5: Histogram of the maximal protrusion $R_{max}$ of amino acids in more than 4,000 high-resolution structures of globular proteins. The 19 amino acids (with glycine being excluded, having no heavy side chain atoms) are denoted with three-letter amino acid code and are colored according to the amino acid classification summarized in Table III. The mean values of $R_{max}$ for each of the amino acids are shown as black X symbols, while the colored rectangles have a width that corresponds to the standard deviation.

We find that there is no significant dependence of $R_{max}$ on the context. However, there are a few cases in which the distributions clearly show resolved multiple peaks. These cases are shown in Figure 6 along with typical conformations that yield the distinct values of $R_{max}$. Except for six amino acids, ILE, GLU, HIS, LYS, and MET (which exhibit more than one peak) and ARG (which has a very broad distribution), the amino acids exhibit one sharp mode in the $R_{max}$ distribution. The most protruding atom in ILE, LYS, MET, and TRP does not depend on the mode, carbon for ILE, MET and TRP and nitrogen for LYS (see Figure 4 for the nomenclature of the atoms in the side chains). For HIS and GLN, the situation is more varied. GLN's lower peak of ~3.8Å in ~70% of cases result from nitrogen atom protrusion while the remaining results from the oxygen atom (see Figure 4). HIS has two close but well-resolved peaks. The more dominant one at ~4.5Å is caused in ~80% of cases by the nitrogen atom protruding the most, while in ~20% of cases the protrude is a carbon atom. In addition, the considerably smaller mode at ~4.7Å is, in more than ~90% of cases, caused by the maximal protrusion of a carbon atom (see Figure 4).

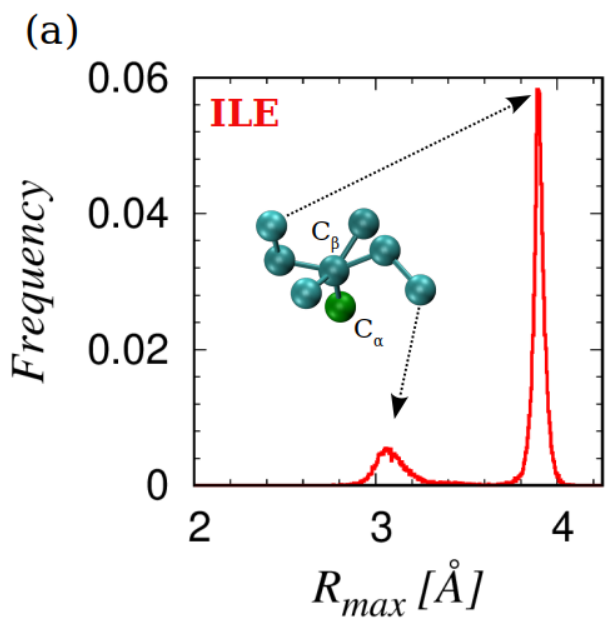
(a) ILE

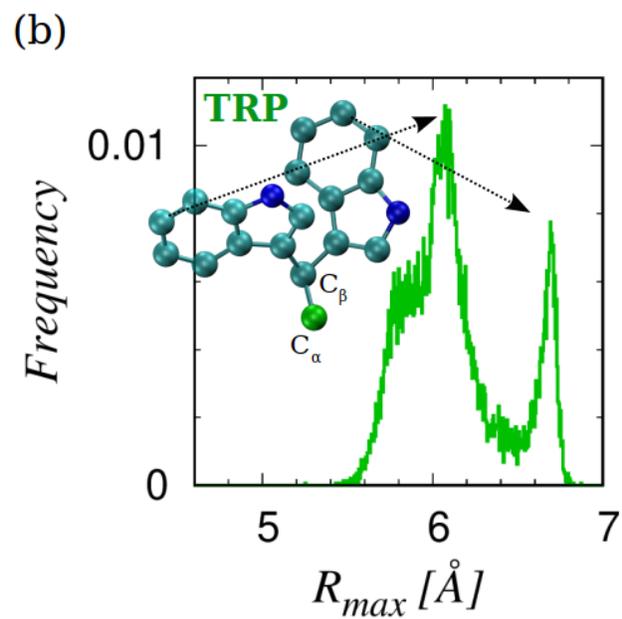
(b) TRP

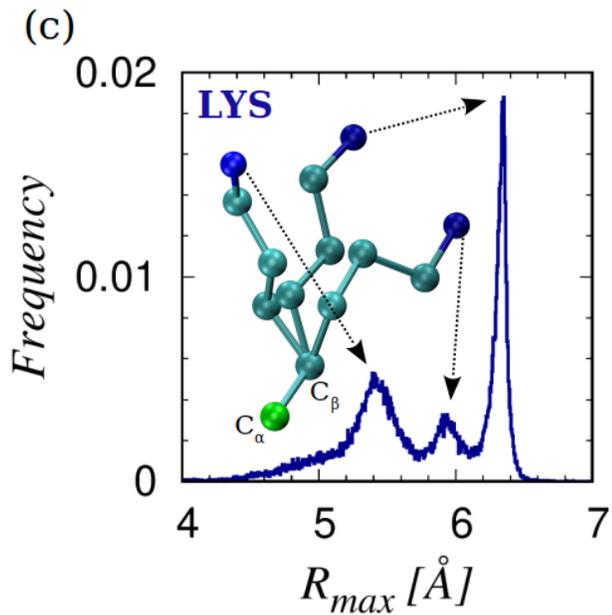
(c) LYS

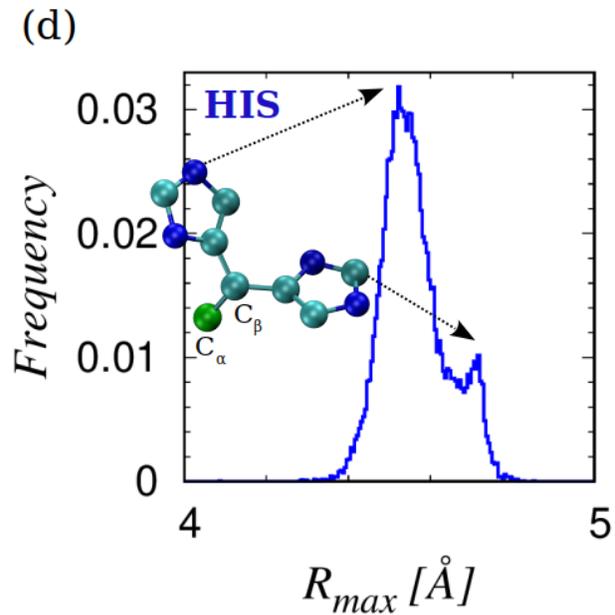
(d) HIS

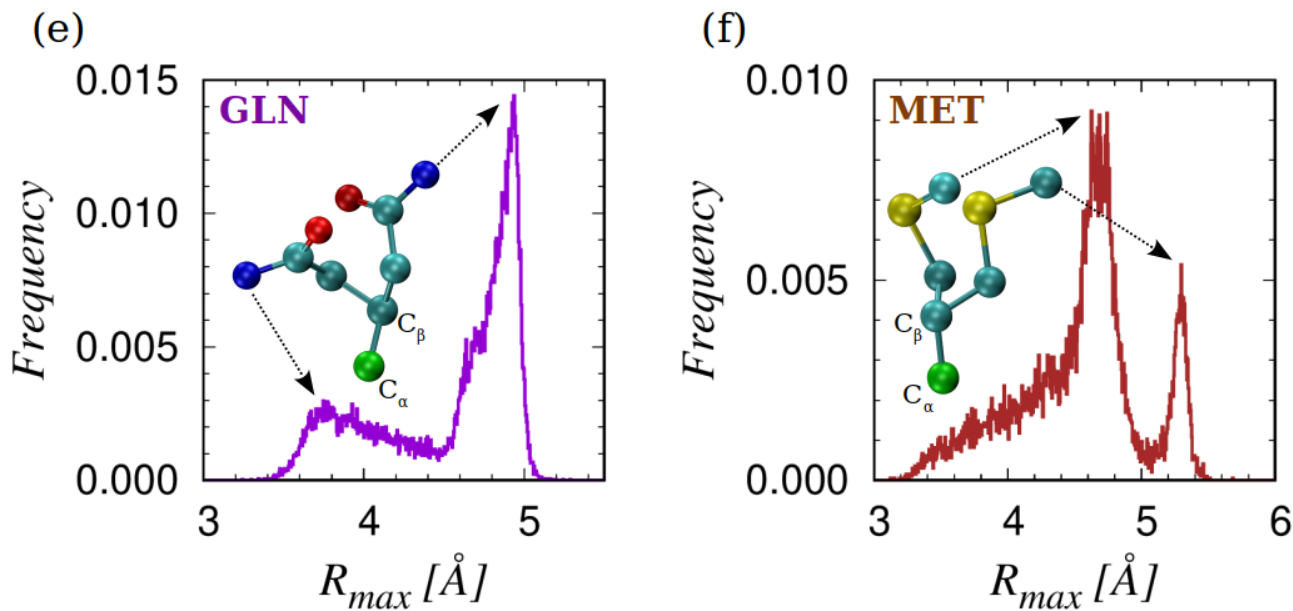

Figure 6: Sketches of the histograms of $R_{max}$ and conformations associated with the multiple modes for six amino acids. For each set of rotamers, the $C_\alpha$ and $C_\beta$ atoms are superimposed to better visualize the distinction between the conformations. The arrows link the maximally protruding atom to the corresponding mode in the $R_{max}$ frequency distribution. The atoms are color coded: carbon $C_\alpha$ in green, carbon C atoms other than $C_\alpha$ in turquoise, oxygen O atoms in red, nitrogen N atoms in dark blue, and sulfur S atoms in yellow.

## IV. CONCLUSIONS

We have presented the results of analyses of the behavior of side chains in experimentally determined native structures of over 4,000 proteins. Our model is simplified, in the spirit of physics, and treats the protein backbone as a chain of $C_\alpha$ atoms. Only the heavy atoms of side chains are considered in our study. To have unbiased standardized results, which allows for variation in bond lengths, we employ a backbone Frenet frame for our analysis.

We have considered several attributes of these side chains. We began with a proxy of structural chemistry by merely considering the constituent heavy atoms in the side chain, the identity of the most protruding atom, and the topology of the side chain (linear or ring) to divide the twenty amino acids into 14 groups. Remarkably, our rudimentary analysis is consistent with careful earlier studies resulting in the development of the much-used IMGT classification [50].

We then turned to the geometry of protrusion and found simplicity in that most side chains lie predominantly in the negative-normal-binormal plane. We went on to analyze the geometry and magnitude of protrusion of the amino acids. Our results show a rich range of behaviors of the side chains in terms of chemistry and geometry. There is a continuum of behaviors with an amino acid for every season.

What we have not considered is the much more formidable many-body problem of side chains in different environments. There are some amino acids that are positioned in in the native state to favorably interact primarily with water molecules and are exposed. Atomistic studies [51-53] considered the interactions between the side chains and different solvents and found that water was optimal, in comparison with cyclohexane (hydrophobic) and ethanol (polar), in the ranges of free energies of solvation. There are other amino acids that are buried in the hydrophobic core by nestling together in a truly many-body manner as a seamless whole.

There are vastly more protein sequences than native state folds and sometimes unrelated sequences adopt the same native state conformation.

Also, native state protein structures are usually robust to significant amino acid mutations [54] except at certain key locations [55-57] in the protein interior, when the side chain packing in the hydrophobic core ceases to remain optimal, so that a sequence switches to a different fold topology that provides a better fit. The amino acid interchangeability is often observed during protein evolution, when amino acid substitutions occur over time, that allows for functional diversification. The availability of a rich repertoire of amino acids facilitates the optimization of sequence within the space of the thousands of putative native state folds.

In future work, we wish to tackle two important questions: What is the detailed role of side chains in selecting the optimal tertiary protein structure? And what are the relative orientations and geometrical and chemical fits of tightly packed side chains that nestle in the protein interior? We plan to pursue a careful study of how one might understand this many-body aspect armed with the insights we have presented here.

**Acknowledgments**

The computer calculations were performed on the Talapas cluster at the University of Oregon.

**Funding information**

This project received funding from the European Union's Horizon 2020 research and innovation program under Marie Skłodowska-Curie Grant Agreement No. 894784 (TŠ). The contents reflect only the authors' view and not the views of the European Commission. JRB was supported by a Knight Chair at the University of Oregon. AG acknowledges support from the Grant PRIN-COFIN 2022JWAF7Y. TXH is supported by The Vietnam Academy of Science and Technology under grant No. NVCC05.02/24-25.

**Author Contributions**

Conceptualization (JRB, TŠ); Data curation (TŠ); Formal analysis (TŠ); Investigation (JRB, TŠ); Methodology (JRB, TŠ), Project administration (JRB, TŠ); Resources (JRB, AG, TXH, AM, TŠ); Software (TŠ); Supervision (JRB); Validation (JRB, TŠ); Visualization (TXH, TŠ); Writing original draft (JRB, TŠ); Writing – review & editing (JRB, AG, TXH, AM, TŠ).

**Conflict of interest**

The authors declare that there is no conflict of interest.